\documentclass[conference]{IEEEtran}
\usepackage[T1]{fontenc}
\usepackage{microtype}
\usepackage{cite}
\usepackage{booktabs}
\usepackage{amsmath,amssymb,amsthm}
\usepackage{xurl}
\usepackage[hidelinks]{hyperref}

\newtheorem{definition}{Definition}
\newtheorem{lemma}{Lemma}
\newtheorem{theorem}{Theorem}
\newtheorem{proposition}{Proposition}
\newtheorem{corollary}{Corollary}

\title{ActWithin: Bounding Wrong-Payment Risk in Delegated Payment Batches}
\author{%
  \IEEEauthorblockN{Shuhao Zheng\IEEEauthorrefmark{1}\IEEEauthorrefmark{2},
    Zhipeng Wang\IEEEauthorrefmark{3}, Zi Xuan Zhang\IEEEauthorrefmark{2},
    and Yang Huang\IEEEauthorrefmark{2}}
  \IEEEauthorblockA{\IEEEauthorrefmark{1}McGill University\quad
    \IEEEauthorrefmark{2}FluxA\quad
    \IEEEauthorrefmark{3}The University of Manchester}
  \IEEEauthorblockA{%
    \href{mailto:shuhao.zheng@mail.mcgill.ca}{shuhao.zheng@mail.mcgill.ca}\quad
    \href{mailto:zhipeng.wang@manchester.ac.uk}{zhipeng.wang@manchester.ac.uk}\quad
    \href{mailto:sky@fluxapay.xyz}{sky@fluxapay.xyz}\quad
    \href{mailto:stella@fluxapay.xyz}{stella@fluxapay.xyz}}
}
\hypersetup{
  pdftitle={ActWithin: Bounding Wrong-Payment Risk in Delegated Payment Batches},
  pdfauthor={Shuhao Zheng, Zhipeng Wang, Zi Xuan Zhang, Yang Huang}
}

\begin{document}
\maketitle

\begin{abstract}
A delegated payment agent can hold valid rights for incompatible revisions of a batch. We study the window in which the selected revision cannot be verified by the execution deadline. For batches that permit every subset of up to \(q\) revisions, ActWithin compiles the limit into a shared quota of \(q\) successful replacement payments. If approved plans are mutually replayable with the same observable inputs, deadlines, and resources, this quota preserves normal completion without additional approvals and minimizes worst-case wrong-payment principal for arbitrary positive amounts. A dynamic program computes its exact risk and witness. For amount-preserving recipient changes, we derive the exact risk of the tightest single amount cap on the same replacement group, provided mixed payment subsets can settle within the window. A two-phase result isolates the risk to payments due before trusted verification. Against the amount cap, an exploratory setting with 16--64 payments and one quarter of positions revisable yields aggregate reductions of 16.23--16.88\% on adapted IBM AMLSim amounts and 10.12--12.52\% on an adapted controlled catalog. The 760 funded original controlled paths with constructed single-position revisions yield 2.69\%. In one local MetaMask fixture, both policies complete all 37 normal plans and the worst wrong debit falls from 93,962 to 81,056 test wei.
\end{abstract}


\section{Introduction}
\label{sec:intro}

In our illustrative scenario, a finance agent is responsible for four payouts of 12, 4, 4, and 4 units. Before leaving the agent unattended, the payer approves the original recipients, one alternative recipient for each payout, and a rule allowing at most one recipient correction. A later business decision selects one plan. The security question arises while the rights for the alternatives are still valid, the payouts are due, and the settlement service cannot yet verify which approved plan was selected.

A compromised agent can use that interval to mix payments from different plans. Every submitted transfer may have the exact recipient, amount, and valid signature required by its credential. Those checks establish that a transfer was included in the approved range; they do not establish that it belongs to the current plan. We ask how much \emph{wrong-payment principal} can be settled, and how far it can be reduced without rejecting any approved normal execution.

\textbf{One batch, two controls.}
Write \(A_i\) for original payment \(i\) and \(B_i\) for its replacement. Suppose the true plan replaces the 12-unit payment, so its intended transfers are \(B_1,A_2,A_3,A_4\). Both controls in Table~\ref{tab:example} retain the same eight exact candidates, 24 units of funding, and a limit of four successful payments. Their only difference is the shared constraint on the four replacements.

\begin{table}[t]
\caption{Running example: amounts \((12,4,4,4)\), at most one revision, and true plan \(B_1,A_2,A_3,A_4\). Both controls complete all five approved plans. Listed effects are the wrong part of an executable attack.}
\label{tab:example}
\centering
\small
\setlength{\tabcolsep}{4pt}
\begin{tabular}{@{}lllr@{}}
\toprule
Replacement control & Limit & Wrong effects & Principal \\
\midrule
Amount cap & 12 units & \(A_1,B_2,B_3,B_4\) & 24 \\
Use quota & 1 use & \(A_1,B_2\) & 16 \\
\bottomrule
\end{tabular}
\end{table}

The smallest shared replacement amount cap is 12, because the approved plan changing the largest payout must complete. That cap also admits all three 4-unit replacements. An attacker can settle the stale original \(A_1\) and those three replacements, debiting 24 units to recipients outside the true plan. A shared quota of one successful replacement blocks this combination. Its worst wrong principal is 16: the stale 12-unit original plus one 4-unit replacement. This remaining risk is unavoidable under the same information. Replaying the complete approved plan \(A_1,B_2,A_3,A_4\) incurs exactly 16 wrong units, yet the executor has no trustworthy evidence with which to distinguish that replay from normal execution.

\textbf{When the window matters.}
A trusted check of the selected version, effective revocation before settlement, or waiting for final approval can resolve this ambiguity when the business deadline permits. Section~\ref{sec:business-window} documents how payment revisions can precede effective revocation. We analyze the remaining risk on payments that cannot wait for trusted evidence. Autonomous agents motivate the contract because choices can be resolved after the user has left and tools may process external, adversarial content~\cite{debenedetti2024agentdojo}. The analysis conditions on a compromised delegate and also applies to conventional automated workers with the same rights and information.

AP2 provides both user-present and autonomous authorization flows~\cite{ap2Spec,ap2Flows}. Its autonomous flow includes safeguards against overlapping mandate presentations. Applying our model therefore requires establishing the specific outstanding-rights and information conditions in a deployment; autonomous operation alone does not imply them. Section~\ref{sec:protocol} identifies a possible composition point and its requirements. The executed implementation evidence in this paper comes from a pinned MetaMask Delegation Framework fixture.

\textbf{Contribution.}
Least privilege motivates retaining required operations while restricting misuse~\cite{saltzer1975protection}. Here the object being restricted is the cumulative combination of payments, with a quantitative objective in principal units. ActWithin keeps the issued alternatives and compiles the approved revision limit into one shared successful-use quota. The mechanism uses established stateful enforcement. Its contribution is the following analysis and evidence.

\begin{itemize}
\item \textbf{An exact risk guarantee.} For the complete family of plans with at most \(q\) revised positions, the quota attains the normal-history replay lower bound, including when original and replacement amounts differ. A dynamic program returns the exact risk and a pair of normal plans witnessing it.
\item \textbf{A matched monetary comparison.} For amount-preserving recipient changes, we derive the exact risk of the tightest single amount cap on the same replacement group, characterize strict improvement and ties, and prove a sharp relative bound. A full-history monitor and a correctly configured manual count quota are matching references.
\item \textbf{Payment-level evidence.} Two amount-source adaptations compare the controls under common resources and approval requirements. A local native execution fixture verifies completion, attack witnesses, quota rejection, and rollback. The evaluation retains small-gain and tie cases alongside the main results.
\end{itemize}

\section{Execution Model and Security Objective}
\label{sec:model}

\subsection{Roles and trust boundaries}
\label{sec:roles}

\textbf{Payer and issuer.} A trusted payer, or a service acting on its approved business policy, defines the payment alternatives and signs their authorization. It also defines which complete payment histories must remain executable. This decision occurs before delegated execution; it can approve the entire revision range in one interaction.

\textbf{Payment agent.} The agent holds the issued rights and submits payment requests. An attacker controls this execution domain, can choose any issued effect, and can reorder, omit, or retry submissions. The attacker can replay a valid candidate from another plan, but cannot forge the payer's signature, change a signed payload, or modify trusted accounting. We condition on this compromise rather than model how often it occurs.

\textbf{Execution point.} One trusted logical execution point authenticates the requested effect, checks shared state, and atomically commits the debit and its accounting updates. It may span several components, provided their combined behavior has these semantics. The model needs neither a gateway per payment nor an independent human approver on every request. Automatic signature and quota checks are part of execution; they do not supply knowledge of the selected business revision.

\subsection{Authorized histories and common resources}
\label{sec:histories}

An authorization epoch begins with a fixed catalog \(E\) of candidate effects, unused rights, and a common funding and resource state. Each effect \(e\) binds an epoch and effect identifier, payer, payee, currency, positive integer principal \(w(e)\), and any additional execution payload. Amounts within an epoch use one currency's smallest unit. Exact-effect deduplication permits each effect to settle at most once. When business-position identifiers already make alternatives mutually exclusive, both compared policies retain that rule.

A history \(h\) records committed effects in order and any execution context visible to enforcement, including timing. Let \(E(h)\) be its set of settled effects. Under a true business plan \(\omega\), the intended effects are \(T_\omega\). The approved contract supplies one or more complete normal histories for each plan, each settling exactly \(T_\omega\). Every admitted normal history must complete by its business deadlines from the same fresh epoch state. These are cumulative successful histories: if the business permits an old payment to settle and then permits an additional amended payment, both belong in the relevant history. A list of mutually exclusive plan snapshots would not describe that contract.

We hold initial funding, amount precision, exact rights, deduplication, global resource caps, deadlines, and trusted interventions fixed across mechanisms. The protected window allows no new approval and no trusted party choosing and executing individual payments on the agent's behalf. Fresh-state completion does not promise recovery after an attacker has already consumed funds, identifiers, or quota.

\subsection{Information and timing}
\label{sec:information}

The protected payments are due before the execution point can verify which approved plan is current. Its view contains the issued catalog and policy, each submitted payload, and committed accounting state. The true plan is used to evaluate correctness; an untrusted claim about that plan is not a trustworthy input to enforcement.

The general lower bound uses \emph{replayable ordered pairs} of plans. A normal history for \(\nu\) is replayable when \(\omega\) is true if the same requests can be submitted within the same window with the same initial enforcement state and resources, authorization validity, timing, and trusted observations. For the sparse-family optimum below, this condition holds between every pair of plans. An authenticated version label, effective revocation, a differing balance, or an earlier deadline can distinguish the executions. If trusted version evidence arrives soon enough to meet the deadline, the executor may use it and the corresponding replay premise no longer holds.

\subsection{Wrong-payment principal}
\label{sec:metric}

\begin{definition}[Wrong principal and policy risk]
\label{def:risk}
For a settled history \(h\) and true plan \(\omega\), define
\begin{equation}
\label{eq:loss}
\ell(h,\omega)=\sum_{e\in E(h)\setminus T_\omega}w(e).
\end{equation}
Let \(\mathcal A_\omega(M)\) be all attack histories that mechanism \(M\) can commit in the protected window from the fixed initial state. Its worst-case risk is
\begin{equation}
\label{eq:risk}
R(M)=\max_\omega\max_{h\in\mathcal A_\omega(M)}\ell(h,\omega).
\end{equation}
\end{definition}

These transfers satisfy their issued credentials but are wrong relative to the selected plan. Each exact effect contributes at most once. The metric counts gross principal of an incorrect exact payment, before recovery, refunds, fees, or compensation; it is not net economic loss. The experiments assign replacement effects payees distinct from their originals. With a same-payee amount correction, crediting the legitimate part would require a different loss function. For randomized mechanisms, normal completion holds with probability one and the risk ranges over possible committed outcomes. The objective is to minimize \(R(M)\) while preserving every admitted timely normal history.

\subsection{The sparse revision family}
\label{sec:envelope}

An epoch has \(n\ge1\) business positions. At position \(i\), the original \(A_i\) has amount \(a_i\) and the replacement \(B_i\) has amount \(b_i\). All \(2n\) candidate effects have distinct identities, keyed by epoch, position, and variant, even when transfer details coincide. Copies of the same authorization retain its effect identity. An optional position identifier can exclude settling both variants of one obligation. The payer approves every complete plan
\begin{equation}
\label{eq:envelope}
T_S=\{A_i:i\notin S\}\cup\{B_i:i\in S\},
\quad S\subseteq[n],\quad |S|\le q.
\end{equation}
Here \(q\) is a nonnegative integer limit supplied by the approved cumulative-history contract and normalized to \(\min(q,n)\). Each plan has an admitted timely history, and all plan pairs meet the replay condition. ActWithin retains the exact rights and existing business deduplication and places one shared successful-use quota on all replacements.

For the amount-preserving comparison, \(a_i=b_i=w_i\), the common funds and total payment cap are \(W=\sum_iw_i\) and \(n\). The compared controls impose no further order or context restriction: any selection of at most one candidate per position can be scheduled within the window, subject to the stated caps. This additional timing condition makes the mixed-plan attack witnesses executable. The general replay lower bound itself does not assume that every mixed selection has a feasible schedule.

\section{Optimal Wrong-Debit Control}
\label{sec:theory}

The analysis has three steps: identify the debit that no timely executor can prevent, attain that value with a shared replacement count, and compare it with the tightest amount cap on the same replacements. Full proofs and the certificate algorithm appear in Appendix~\ref{sec:proofs}.

\subsection{A lower bound from normal-history replay}

Suppose the executor must accept a normal history for plan \(\nu\), and the same history can be presented when \(\omega\) is the true plan. If its observable inputs, authorization validity, resources, and timing are unchanged, the executor must make the same decisions. The wrong principal is then the amount of payments in \(T_\nu\) that are absent from \(T_\omega\). Under the full pairwise replay condition of Section~\ref{sec:model}, define
\begin{equation}
\label{eq:diameter}
D_{\mathcal F}=\max_{\omega,\nu}
    \sum_{e\in T_\nu\setminus T_\omega}w(e).
\end{equation}
The difference is directed: replaying \(\nu\) in \(\omega\) need not cost as much as replaying \(\omega\) in \(\nu\).

\begin{lemma}[Normal-history replay]
\label{lem:replay}
With common fresh resources and full pairwise replay before the payment deadlines, every policy preserving all admitted normal histories has \(R(M)\ge D_{\mathcal F}\). A monitor that accepts exactly observable prefixes of admitted normal histories attains equality.
\end{lemma}

The lower bound follows because timely completion forces the executor to accept the replay. For the matching upper bound, every history accepted by the full monitor is contained in an admitted plan; its wrong principal cannot exceed that plan's directed difference. If only some ordered plan pairs meet the replay conditions, only those pairs may enter the lower-bound maximum. Equality with that restricted bound requires a separate argument. The results below use full pairwise replay. A trusted signal identifying the selected plan in time changes this premise.

\subsection{One count quota attains the bound}

ActWithin retains the exact permissions and limits successful settlements from \(\{B_1,\ldots,B_n\}\) to \(q\) during the epoch. In a true plan \(T_S\), a wrong original must be at a replaced position in \(S\); a wrong replacement must be outside \(S\). There can be at most \(q\) of each: the first limit follows from the approved plan, and the second from the shared counter.

\begin{theorem}[Optimal risk for sparse amendments]
\label{thm:minimax}
Assume the complete family in \eqref{eq:envelope}, distinct identities for all \(2n\) effects, positive amounts, fresh common resources supporting every admitted history, and full pairwise replay within the execution window. The shared replacement count preserves every admitted normal history and satisfies
\begin{equation}
\label{eq:dq}
\begin{aligned}
R_{\mathrm{count}}&=\min_M R(M)=D_q,\\
D_q&=\max_{\substack{I,J\subseteq[n],\ I\cap J=\varnothing\\
                     |I|\le q,\ |J|\le q}}
     \left(\sum_{i\in I}a_i+\sum_{j\in J}b_j\right),
\end{aligned}
\end{equation}
where the minimum is over policies satisfying the same information and completion contract. This holds with exact-effect deduplication alone or with additional position-level deduplication.
\end{theorem}

\textit{Why it holds.} Every normal history contains at most \(q\) replacements, in any admitted order. For an attack, let \(I\) index its wrong originals and \(J\) its wrong replacements. These sets are disjoint and satisfy the two cardinality bounds in \eqref{eq:dq}, proving the upper bound. For any maximizing pair, replay the normal plan \(T_J\) when \(T_I\) is true. Its wrong effects are exactly \(\{A_i:i\in I\}\cup\{B_j:j\in J\}\), so Lemma~\ref{lem:replay} supplies the matching lower bound.

This witness is a complete normal plan and needs only normal resources. With debit-only funding, the largest normal outlay is
\begin{equation}
\label{eq:funding}
C=\sum_i a_i+\operatorname{Top}_q\big((b_i-a_i)_+\big),
\end{equation}
where \(\operatorname{Top}_k\) sums the \(k\) largest entries and \((x)_+=\max(x,0)\). A common balance of \(C\) suffices for every plan. Other resource and timing restrictions must also permit the replay.

The compiler computes \(D_q\) and a maximizing pair by a dynamic program that assigns each position to \(I\), to \(J\), or to neither. The value computation uses \(O(n(q+1)^2)\) arithmetic operations and \(O((q+1)^2)\) rolling space; storing predecessors uses \(O(n(q+1)^2)\) space to recover a witness. Appendix~\ref{sec:certificate-dp} gives the recurrence and verification procedure. For the four-payment example, \(I=\{1\}\) and \(J=\{2\}\) give \mbox{\(D_1=12+4=16\)}. Replaying \(T_{\{2\}}\) settles four payments totaling 24 units, of which 16 are wrong under \(T_{\{1\}}\).

\subsection{Exact comparison with one shared amount cap}

Now suppose each position's two variants have equal amounts, \(a_i=b_i=w_i\), while amounts may differ across positions. Every normal plan spends \(W=\sum_iw_i\) in \(n\) payments. Both policies have funding and a global spend cap of \(W\), a global count cap of \(n\), the same exact permissions and deduplication, and the same deadlines. For this comparison, the remaining checks allow any subset with at most one effect per position to be scheduled within the window; there are no additional sequence or context guards. Each policy adds one shared cap on the identical replacement group.

Let \(Q\) contain the \(q\) largest amounts, with ties broken arbitrarily, and write \(w(J)=\sum_{j\in J}w_j\). Define
\begin{align}
H_q &= w(Q), \nonumber\\
K_q &= \max\{w(J):J\subseteq[n]\setminus Q,\ w(J)\le H_q\},\nonumber\\
S_q &= \operatorname{Top}_{\min(q,n-q)}\big((w_i)_{i\notin Q}\big).
\label{eq:amount-terms}
\end{align}
Thus \(K_q\) is the largest remaining subset that fits an amount budget of \(H_q\), whereas \(S_q\) is the largest remaining subset using at most \(q\) positions.

\begin{theorem}[Tightest matched amount cap]
\label{thm:amount}
Under these matched controls and the replay conditions above, the smallest replacement amount cap preserving all normal plans is \(H_q\), and
\begin{equation}
\label{eq:matched-risks}
R_{\mathrm{money}}=H_q+K_q,\qquad
R_{\mathrm{count}}=H_q+S_q.
\end{equation}
The count policy strictly improves risk if and only if \(K_q>S_q\), by exactly \(K_q-S_q\) units.
\end{theorem}

\textit{Why it holds.} Replacing the positions in \(Q\) is an approved plan, so the amount cap must accommodate \(H_q\). In the worst attack, an exchange argument places the wrong originals at those positions. The remaining wrong replacements can use any disjoint subset fitting the cap, yielding \(K_q\). A count cap limits this second set to \(q\) positions, yielding \(S_q\). The money-policy witness executes \(\{A_i:i\in Q\}\) and the maximizing remaining replacements while \(T_Q\) is true. It uses at most \(n\) payments and \(W\) funds. The scheduling condition makes this mixed witness executable by the same deadline.

In the running example, \(H_1=12\). All three 4-unit replacements fit that amount cap, so \(K_1=12\), whereas \(S_1=4\). The risks are therefore 24 and 16. The policies tie at zero when \(q=0\), at \(W\) when \(q\ge n/2\), and for every \(q\) when all position amounts are equal. With additional ordering restrictions, the formulas remain upper bounds; exactness of the monetary comparison requires checking that its witness remains feasible.

\begin{corollary}[Sharp relative reduction]
\label{cor:sharp-gap}
For \(n>2q\ge2\),
\begin{equation}
\label{eq:sharp-gap}
\frac{R_{\mathrm{money}}-R_{\mathrm{count}}}{R_{\mathrm{money}}}
\le\frac{n-2q}{2(n-q)}.
\end{equation}
Equality is attained by \(q\) amounts of \(n-q\) and \(n-q\) amounts of \(q\).
\end{corollary}

The largest \(q\) remaining amounts hold at least a \(q/(n-q)\) share of the remaining total, while \(K_q\le H_q\); these two inequalities give the bound. At \(n=4q\), the sharp ceiling is one third. This is a worst-case separation bound, not an empirical average.

\paragraph{Comparison boundary}
A correctly configured manual count and a full normal-history monitor attain the same risk as ActWithin. Multiple overlapping amount constraints can also encode the count limit, as shown in Appendix~\ref{sec:representation}. The comparison isolates one shared amount cap on the same group. The appendix also states the separate, representation-dependent distinction between attaining optimal risk and representing every legal effect set.

\subsection{Payments That Can Wait for Verified Revision Evidence}
\label{sec:deadline-split}

Consider a fixed schedule with two phases. A set \(U\) of positions must settle before trusted plan evidence arrives. Enforcement admits only positions in \(U\) in the first phase. All remaining positions can wait for mandatory verification and still meet their deadlines. Assume the plan is selected before the first phase, the early normal prefixes are replayable with common resources, and the phase order supports every admitted plan. In the second phase, both policies accept only verified effects at the remaining positions. Let \(m=|U|\) and \(r=\min(q,m)\).

\begin{proposition}[Risk with partial waiting]
\label{prop:deadline-split}
For amount-preserving revisions, the minimum worst-case wrong principal is \(\operatorname{Top}_{\min(2q,m)}((w_i)_{i\in U})\). A shared early quota of \(r\) attains it. If mixed subsets of early positions can settle by their deadlines, the tight shared amount cap over the early replacements gives \(H_r+K_r\) from \eqref{eq:amount-terms}, computed on \(U\).
\end{proposition}

A trusted check on all positions before their respective deadlines gives zero wrong principal under the model. When \(U\) is nonempty, requiring every position to wait for the same late evidence misses an early deadline. The proposition keeps the original money-versus-count comparison for the positions that must be released first. A proof appears in Appendix~\ref{sec:deadline-proof}.

\section{Application and Enforcement}
\label{sec:protocol}

\subsection{From an Approved Batch to Settlement}
\label{sec:enforcement-lifecycle}

The payer or its trusted issuing service supplies the batch contract: an epoch identifier, obligation positions, exact original and replacement effects, and the permitted cumulative amendment limit \(q\). Each effect fixes the funding account, currency, recipient, amount, and payment reference. A verified recipient directory or an approved substitution rule can supply the alternatives. The agent may propose them, but the trusted issuer determines which effects enter the signed contract and which consume the replacement allowance. The contract must cover all permitted successful payments over the epoch, including any amendments allowed after earlier payments have settled. Inferring that contract from logs or natural-language instructions is a separate, unvalidated step.

The execution path has four stages.
\begin{enumerate}
\item \textbf{Issue.} The issuer approves the complete family of plans, signs the exact effects and their group membership, and installs the common resource limits and a replacement count initially equal to zero. The compiler supplies the risk value and a pair of normal plans witnessing it.
\item \textbf{Check.} On an agent submission, one logical trusted execution point verifies the signature, epoch, exact payload, validity interval, effect identifier, and any position-level exclusion. It also checks the common funds and payment count and, for a replacement, that another successful use fits within \(q\).
\item \textbf{Settle and commit.} The executor serializes competing requests against the same state. It provisionally reserves the required resources, performs the authorized transfer, and commits the debit, identifiers, and counters as one atomic transition. A committed replacement consumes one group unit; an original consumes none.
\item \textbf{Roll back on failure.} A failed check or transfer leaves no debit and no committed consumption of an identifier, global allowance, or replacement allowance. Retries therefore observe the last committed state.
\end{enumerate}
These are automated checks at the settlement boundary. They require neither a new user approval nor a trusted operator to choose or execute each payment on the agent's behalf. A physical deployment may distribute components, provided it implements this single logical transition.

\paragraph{Preserving an epoch across reissuance}
All still-executable credentials for one epoch must charge the same authoritative state. Re-signing a credential must not replenish its allowance. A deployment can retain a stable epoch ledger across credential generations, or start a new epoch only after old authority can no longer settle. Changing the catalog, \(q\), or the permitted sequence of revisions requires rechecking the cumulative history contract. Merely changing a credential hash does not satisfy this requirement. The local fixture below uses one fixed signed policy per arm; it does not test migration across reissued credentials.

\subsection{Documented Revocation Delay and the Execution Window}
\label{sec:business-window}

Payment systems can separate a revision from its enforcement. CheckIssuing documents payee corrections by reissue. After the platform cancels the original, a mailed check or a viewed digital check can still be deposited unless the bank applies a separate stop~\cite{checkIssuingReissue}. A replacement that clears approval after the cutoff may miss the current mailing cycle~\cite{checkIssuingCutoff}. In a separate audit of CUNY, the New York State Comptroller found that delayed reconciliation led to a duplicate payment of \$544. The original check was cashed before the stop was placed, and the payment was never recovered~\cite{nyCunyDisbursements}. New York State's accounting guidance and the Social Security Administration's reissue procedures likewise treat confirmation of the original check's status as a distinct decision point~\cite{nyPaidOverStop,ssaDoubleChecks}.

These workflows illustrate separate timing for revision, verification, and revocation. CheckIssuing reissues checks sequentially and applies its ordinary approval process to each replacement. Our hypothetical batch adds concurrently executable exact alternatives, common settlement accounting, and a preapproved revision limit. Plaid's funded-ledger payout flow supplies another example of staged authorization. Plaid publishes same-day ACH submission cutoffs and separately specifies that cancellation ends at network submission~\cite{plaidVendorPayouts,plaidCreatingTransfers}.

An issuer can carry independent, authenticated plan evidence through the agent and require it at settlement. For an epoch with a single binding plan decision, an epoch-bound issuer signature can suffice. When the issuer can sign successive revisions, the executor also needs freshness evidence. A signature alone does not establish latestness. If presentation is optional and the executor allows a no-proof fallback, the agent can omit the proof. A deployment should determine which positions can wait for mandatory verification or effective revocation while preserving their payment deadlines. Section~\ref{sec:deadline-split} analyzes the positions that cannot wait.

\subsection{Relationship to AP2}
\label{sec:ap2}

We use AP2 v0.2 as documented on November 20, 2026~\cite{ap2Spec}. A Checkout Mandate authorizes completion of a merchant checkout; its closed form contains the merchant-signed checkout. A Payment Mandate authorizes payment and binds the payee, amount, instrument, and checkout reference~\cite{ap2Checkout,ap2Payment}. In the human-present flow, a Trusted Surface obtains consent for the closed mandates. In the autonomous flow, the user approves open mandates containing constraints and binding the agent's key; the agent later presents these together with its signed closed mandates~\cite{ap2Flows}. The Merchant verifies checkout authorization, the Credential Provider and applicable Network verify payment authorization, and the Merchant Payment Processor verifies the payment credential's scope~\cite{ap2Spec}. These are verification roles, which may be combined within one entity.

AP2 already specifies cumulative budgets and recurrence limits, including a maximum occurrence count~\cite{ap2Payment}. Its autonomous flow restricts overlapping mandate creation pending an error receipt, and implementation guidance describes deterministic checks around the agent's signing key to prevent overlaps~\cite{ap2Flows,ap2Implementation}. The mandate schema version identifies the credential format~\cite{ap2Spec}. Our use of revision refers to the current set of approved business payments. If protected signing or timely plan verification prevents the relevant alternatives from remaining executable, the replay premise of Section~\ref{sec:model} does not hold.

A possible composition would add an authenticated mapping from effects to obligation positions, epochs, and replacement groups at an executor that can maintain shared settlement state. It must preserve AP2's signature, binding, constraint, and receipt checks. AP2's approval and presentation accounting also needs an explicit mapping to the successful-settlement accounting used here~\cite{ap2Payment}. The current Checkout Mandate line-item constraint does not itself split one open mandate across multiple checkouts~\cite{ap2Checkout}. Thus a batch rule requires application-level semantics and an identified enforcement path. The contribution is the exact risk analysis for that rule under the stated information contract. We have not implemented or measured an AP2 batch integration.

\subsection{Realization with Existing MetaMask Enforcers}
\label{sec:metamask}

The local conformance experiment uses MetaMask Delegation Framework commit \nolinkurl{b4b8c41fc912455a1af2300bfcd1276676b92a2e}~\cite{metamaskFramework}. In the fixture, the issuer signs a disjunction of exact transfers. The \nolinkurl{LogicalOrWrapperEnforcer} selects a branch, \nolinkurl{ExactExecutionEnforcer} fixes its recipient, value, and calldata, and \nolinkurl{IdEnforcer} prevents reuse of its business position. Replacement branches additionally invoke one shared group guard: \nolinkurl{LimitedCallsEnforcer} for ActWithin or \nolinkurl{NativeTokenTransferAmountEnforcer} for the monetary comparator. At this pinned version, those group calls use the same wrapper caller and delegation hash, producing a common counter across replacement branches. Separate guards impose the same global count and spending limits in both arms.

The delegated agent submits through the public redemption entry point. At the pinned revision, revocation updates the on-chain disabled state for a delegation hash, and redemption checks that state~\cite{metamaskManager}. In the fixture both variants share one signed delegation per arm, so disabling that delegation also removes the current plan's route. Timely revocation preserves normal completion only if the chosen effects remain executable through valid authority. The local tests check rejected-request rollback for the tested execution path (Section~\ref{sec:native-evaluation}). The successful-settlement interpretation requires atomic accounting. Shared balances across other batches, bypass routes, and execution modes with different failure semantics need separate conformance checks.

\section{Evaluation}
\label{sec:evaluation}

The evaluation asks how much a shared replacement count lowers worst-case wrong-payment principal relative to the tightest single spending cap on the same replacements. We compute exact risks on two amount adaptations, retain the original-path sensitivity, and examine a matched local execution through the MetaMask Delegation Framework. These provide evidence about amount composition, the specified authorization contract, and executable enforcement, respectively.

\subsection{Amount Sources and Construction}

\textbf{AMLSim.} We use all 356,613 simulated transaction amounts in the three official sample archives at the pinned IBM AMLSim revision~\cite{amlSim}. Within each archive, rows retain their archived order and are partitioned into consecutive, nonoverlapping batches of a fixed width. Amounts are positive and represented exactly in hundredths of a simulated unit. Recipients, payment rights, and revisions are constructed for this experiment; the original transaction graph and anti-money-laundering task are not evaluated.

\textbf{Controlled catalog.} We use all 1,980 action amounts from two existing controlled payment scenario files, contributing 820 and 1,160 amounts. Each amount is taken once in case and action-array order, without scaling or selection by outcome. Each file is batched independently. Amounts retain the source records' integer-cent units. The adaptation constructs new obligations, accounts, recipients, and authorizations, with a balance equal to each batch's total. It does not merge the original accounts or preserve their workflow dependencies. The original paths and balances are evaluated separately in Appendix~\ref{sec:original-paths}.

\begin{table}[t]
\centering
\caption{Reduction in aggregate worst-case wrong-payment principal against the tight shared replacement spending cap, for \(q=n/4\). Each percentage uses the ratio of risk sums in \eqref{eq:aggregate-reduction}; its batch denominator is shown.}
\label{tab:quarter-risk}
\setlength{\tabcolsep}{4pt}
\begin{tabular}{@{}rr rr rr@{}}
\toprule
&& \multicolumn{2}{c}{AMLSim} & \multicolumn{2}{c}{Controlled catalog} \\
\cmidrule(lr){3-4}\cmidrule(l){5-6}
\(n\) & \(q\) & Batches & Reduction & Batches & Reduction \\
\midrule
4  & 1  & 89,152 & 8.20\%  & 495 & 2.88\%  \\
8  & 2  & 44,575 & 14.38\% & 247 & 8.40\%  \\
16 & 4  & 22,286 & 16.23\% & 123 & 10.12\% \\
32 & 8  & 11,143 & 16.71\% & 61  & 12.18\% \\
64 & 16 & 5,570  & 16.88\% & 30  & 12.52\% \\
\bottomrule
\end{tabular}
\end{table}

Both adaptations use widths \(2,4,8,16,32,64\). At each width, quotas are the distinct values in
\[
\{1,\max(1,\lfloor n/8\rfloor),\max(1,\lfloor n/4\rfloor),
  \lfloor n/2\rfloor,n\}.
\]
Every complete batch is evaluated at every quota for that width. Incomplete final batches are recorded and excluded from risk statistics; Appendix~\ref{sec:evalmore} reports their counts and all 24 width--quota strata per corpus. The constructions impose no additional source-workflow order or context guard: the candidates are pending within a common execution window that permits the witness schedules and any sequence of at most \(n\) settlements.

\subsection{Matched Policies and Measurement}
\label{sec:eval-contract}

At position \(i\), the original \(A_i\) and replacement \(B_i\) have the same amount \(w_i\) and distinct payees. The normal family contains every complete plan with at most \(q\) replacements. Both policies start from fresh state with identical exact one-use rights, precision, a global successful-payment cap of \(n\), and funding and total-spend cap \(W=\sum_i w_i\). Both preserve the same deadlines and admit every normal plan. Neither requires a further approval or a trusted operator to choose and submit an individual payment. Automated checks by the trusted executor remain active in both arms.

The monetary policy puts all \(B_i\) in one shared spending bucket of size \(H_q\), the sum of the \(q\) largest amounts. This is the smallest such cap that preserves every normal plan. ActWithin applies a shared limit of \(q\) successful settlements to exactly the same \(B_i\). The catalog runs use distinct one-use IDs for the variants. Adding a shared one-use business ID at each position leaves both exact risks unchanged: their maximizing witnesses need at most one effect per position. The native fixture uses these position IDs in both arms.

For the complete batches \(\mathcal B\) in one corpus, width, and quota stratum, we report
\begin{equation}
\label{eq:aggregate-reduction}
\Delta(\mathcal B)=
\frac{\sum_{b\in\mathcal B} R_{\mathrm{money}}(b)
      -\sum_{b\in\mathcal B} R_{\mathrm{count}}(b)}
     {\sum_{b\in\mathcal B} R_{\mathrm{money}}(b)}.
\end{equation}
This is a reduction in aggregate worst-case principal, with larger baseline risks receiving greater weight. It is not the mean of per-batch percentages or an observed attack rate. Exact integer subset-sum computes the monetary optimum, and the count optimum follows from Theorem~\ref{thm:amount}. Appendix~\ref{sec:eval-validation} describes the recorded independent checks. A full normal-history monitor and a correctly configured manual shared count tie ActWithin's risk; they are matching references. Removing the replacement guard while retaining the common global caps gives a weaker ablation.

\subsection{Results and Sensitivity}

At \(q=n/4\), Table~\ref{tab:quarter-risk} shows reductions of 16.23--16.88\% on AMLSim and 10.12--12.52\% on the controlled catalog at widths 16--64. Every AMLSim batch at these three widths improves. In the controlled catalog, 121 of 123 batches improve at width 16, the remaining two tie, and every batch improves at widths 32 and 64. Removing each AMLSim archive in turn gives ranges of 16.21--16.25\%, 16.67--16.73\%, and 16.84--16.92\%, respectively. Each controlled source also improves when evaluated alone. Appendix~\ref{sec:evalmore} gives the risk sums, source sensitivity, and all other strata.

Gains are smaller under other contracts. With only one allowed replacement, the AMLSim reductions at widths 32 and 64 are 2.62\% and 1.32\%. Every batch in both adaptations ties when \(q\ge n/2\), as the theory predicts. Keeping the original controlled paths gives a 2.69\% aggregate reduction at \(q=1\); the largest reduction in that path grid is 8.49\% at \((n,q)=(8,2)\). Thus the larger catalog gains depend on the constructed batch composition and permitted revisions, rather than establishing one improvement level for all payment workflows.

The study is exploratory. After earlier results under the stronger comparator showed small gains, the experiments expanded the quota and width grids. Each extension fixed its protocol before execution and reports its complete grid, but there is no independently isolated final holdout. Widths and quotas reuse source amounts, and adjacent amounts or paths may be dependent. The source-removal ranges describe corpus-composition sensitivity, not population confidence intervals. These results concern finite adapted corpora under the stated information and resource contract.

\subsection{Native Execution}
\label{sec:native-evaluation}

We compare both guards through the public redemption entry point of MetaMask Delegation Framework commit \texttt{b4b8c41f}~\cite{metamaskFramework}. The predetermined fixture uses the first eight rows of the lexicographically first AMLSim archive, with \(q=2\). Both arms use the same signed delegation topology, exact execution branches, position IDs, and global guards. Only the shared replacement guard differs: a native transfer-amount cap of 47,899 test wei or a successful-call cap of two. The source integers are interpreted as test wei directly, without currency conversion.

From fresh state, each arm completes all \(1+8+\binom{8}{2}=37\) normal plans. Every plan executes eight payments in source-position order and debits the full initial balance of 125,233 test wei. Post-issuance approvals and trusted root execution calls are zero. The shared count lowers the executed worst-case wrong debit from 93,962 to 81,056 test wei, a reduction of 12,906 test wei (13.74\%).

\begin{table}[t]
\centering
\caption{Native wrong-payment witnesses. Positions are numbered from one. The true plan replaces positions 6 and 7; both attacks execute their stale originals for 47,899 test wei. All listed replacement payments are also wrong for that plan.}
\label{tab:native-witness}
\footnotesize
\setlength{\tabcolsep}{3.5pt}
\begin{tabular}{@{}lcrr@{}}
\toprule
Guard & Wrong \(B_i\) positions & \(B_i\) wei & Total wei \\
\midrule
Amount & \(1,2,4,8\) & 46,063 & 93,962 \\
Count  & \(1,3\)     & 33,157 & 81,056 \\
\bottomrule
\end{tabular}
\end{table}

Table~\ref{tab:native-witness} shows that four small replacements fit the monetary bucket and only two fit the shared count. The recorded tests also reject an additional replacement under each guard and verify that a rejected branch or exact-payload check rolls back balance and global, replacement, and position-ID state. All four native tests pass. The true revision supplies the evaluation label, and the executor runs without that label. This local fixture exercises the 37 plans in source-position order. The deadline and arbitrary admitted orders are conditions of the formal analysis. Appendix~\ref{sec:eval-artifact} states the evidence and reproduction boundary.

\subsection{Sensitivity to Waiting for Verification}
\label{sec:deadline-evaluation}

This exact sensitivity calculation reuses the same eight payment amounts and \(q=2\) and enumerates all \(2^8\) early-position sets. The model lets both quota controls satisfy the constructed deadlines and requires verified effects on all later positions. Requiring the same late proof for every payment meets all deadlines only for the empty early set. We exclude the other wait-only cases from loss comparisons at matched completion.

\begin{table}[t]
\centering
\caption{Exact sensitivity on one fixed eight-payment catalog. For each early width, the reduction pools risk sums across all subsets of that size. Widths 1--4 tie individually and are grouped here.}
\label{tab:deadline-split}
\footnotesize
\begin{tabular}{@{}rrrr@{}}
\toprule
Early positions & Subsets & Count strictly lower & Reduction \\
\midrule
0 & 1 & 0 & -- \\
1--4 & 162 & 0 & 0.00\% \\
5 & 56 & 51 & 12.31\% \\
6 & 28 & 28 & 13.49\% \\
7 & 8 & 8 & 14.37\% \\
8 & 1 & 1 & 13.74\% \\
\bottomrule
\end{tabular}
\end{table}

The policies tie for every subset with at most four early positions and have zero risk when all payments can wait. The count reduces risk for 88 of the 93 larger early sets. The calculation agrees with a separate exhaustive enumeration of 65,536 classifications and with the frozen all-early native result. These constructed schedules reuse the same amounts and measure sensitivity to which payments are due before verification.

\section{Related Work}
\label{sec:related}

\textbf{History-sensitive authorization.}
Least privilege and separation of privilege are established design principles~\cite{saltzer1975protection}. Consumable credentials and history-based capabilities make the exercise of a right change future authorization~\cite{bowers2007consumable,tandon2018hcap}. Security automata provide the standard execution-monitor perspective~\cite{schneider2000enforceable}; our full-history reference is an instance of it. Workflow authorization connects prior actions, separation of duties, and the ability to finish a workflow~\cite{basin2009dynamic,basin2012workflow}. Role separation supplies the trusted issuer and enforcement boundary in our model. Obtaining a further trusted decision about the selected plan can improve risk when it is available by the deadline. Within a fixed information and completion contract, our result quantifies the exact principal that stateful enforcement can protect.

\textbf{Constraint synthesis and representation.}
Petri-net region synthesis derives places from specified behavior~\cite{bergenthum2007regions}. A shared replacement quota has the familiar semantics of a place with \(q\) initial tokens, each consumed by one replacement. The dynamic program and subset-sum computation likewise use standard techniques. Our analysis identifies the minimax payment risk for this revision family and the exact penalty of replacing its use quota with one monetary bucket. Appendix~\ref{sec:proofs} also distinguishes the logical constraints needed to attain optimal risk from those needed to represent all legal effect sets, under an explicitly restricted representation.

\textbf{Payment mandates and agent execution.}
AP2 specifies linked checkout and payment authorization, including open constraints, recurrence counts, and budgets~\cite{ap2Spec,ap2Payment}. These existing controls establish the relevant protocol context. Our proposed mapping needs authenticated batch membership, settlement-consistent accounting, and the stated information window. The MetaMask Delegation Framework provides exact execution and stateful enforcers~\cite{metamaskFramework}, which implement both controls in our native fixture. AgentDojo evaluates task behavior under prompt injection~\cite{debenedetti2024agentdojo}. It motivates compromise of an agent's execution domain; our payment experiments use separate amount sources and measure risk after compromise.

\section{Scope, Deployment Choices, and Limitations}
\label{sec:limits}

\textbf{Choosing a control.}
Enforce the selected plan wherever a trusted version check fits the payment deadline. Under the two-phase schedule of Proposition~\ref{prop:deadline-split}, a quota covers the positions that must settle first. For amount-preserving replacements, \(K_r>S_r\) identifies extra risk under a single replacement amount cap when mixed early payments can settle by their deadlines. Equal amounts and \(r\ge |U|/2\) give ties under this comparison. A full-history monitor or an existing correctly scoped count quota already reaches the same floor. Multiple overlapping amount constraints can also express the quota.

\textbf{Inputs and shared state.}
The issuer must supply trusted business positions, exact alternatives, and the set of permitted cumulative histories. Extracting those inputs automatically from logs or natural language is unvalidated. Other revision patterns require a different model. All redemption routes must share the relevant epoch accounting. Cross-batch balances, overlapping policies, and reissued credentials can change feasible attack and normal histories, so their combined resource state must be analyzed. The local fixture establishes the stated single-batch behavior; it does not measure migration cost, cross-issuer coordination, or production latency.

\textbf{Availability and economic interpretation.}
An attacker can consume quota or a position identifier before an honest request. Our completion guarantee starts from unused state and does not guarantee attack recovery or uninterrupted service afterward. The objective measures gross wrong principal, without estimating compromise frequency, recoverability, reversals, fees, or net economic loss. Trusted issuer, executor, and accounting integrity remain assumptions.

\textbf{External validity.}
The amount adaptations construct recipients, authorizations, and revision sets. Reusing source amounts across widths and quotas does not create independent samples of real business amendments. The retained original-path experiments show that gains can be small. The native result covers one predetermined local fixture, and normal plans were executed in source position order. Arbitrary admitted order is a formal guarantee, not an exhaustive execution measurement. Deployment prevalence, window duration, real revision distributions, live attack outcomes, and AP2 compatibility require additional evidence.

\textbf{Data and research conduct.}
The public amount source is simulated, the second source is a controlled scenario catalog, and the native tests use local test accounts without live transfers. Appendix~\ref{sec:eval-artifact} identifies the supplied records and the additional archives, runners, and native test patch needed for a complete rerun. We make no additional claim about public release of the controlled source files.

\section{Conclusion}
\label{sec:conclusion}

When alternative payments can settle before the executor can verify the selected plan, preserving normal execution imposes a replay-based floor on wrong-payment principal. For the complete family of at most \(q\) revisions, one shared replacement-use quota reaches that floor. A matched monetary cap may admit more small replacements than the approved revision count. The formulas identify both strict gains and natural ties, while the corpus and native results quantify that difference under their stated contracts. Applying the result requires validating the information window, the cumulative-history model, and common settlement state.

\section*{Data and Code Availability}

The references pin the public AMLSim archives and MetaMask revision. Appendix~\ref{sec:eval-artifact} gives the eight amounts and quota for the deadline calculation. This release includes the paper and LaTeX sources. Sharing controlled inputs and derived records requires permission.

\section*{LLM usage considerations}

Large language models assisted with literature discovery, development and checking of mathematical arguments, experiment-code scaffolding, and manuscript drafting and editing. The risk calculations use deterministic computation, and the conformance results come from local execution.

\section*{Ethical Considerations}

The evaluation uses public simulated amounts, a controlled payment catalog, and local test funds. The bound covers gross payment principal under the stated enforcement and information assumptions.

\bibliographystyle{IEEEtran}
\bibliography{references}

\appendices
\section{Proofs and Risk Certificates}
\label{sec:proofs}

\subsection{Normal-history replay (Lemma~\ref{lem:replay})}

Fix an ordered pair \((\omega,\nu)\) satisfying the replay condition and an admitted complete history for \(\nu\). Present the same requests, timing, and observable context while \(\omega\) is the true plan. The initial enforcement view and resources coincide. Inductively, identical visible requests and prior decisions induce identical settlement and accounting transitions. A policy preserving timely normal completion must therefore accept the replay, whose wrong principal is
\(\sum_{e\in T_\nu\setminus T_\omega}w(e)\).
For a randomized policy, couple its random choices in the two executions. Their observable distributions coincide, and probability-one normal completion implies probability-one completion of the replay. Randomization cannot lower the worst-case risk. Taking the maximum over replayable pairs proves the lower bound. Under full pairwise replay this is \eqref{eq:diameter}; otherwise the maximum must be restricted.

For attainability under full pairwise replay, the full-history monitor retains the observed sequence and accepts a settlement only if the extended sequence, including its observable context, is a prefix of an admitted normal history. Every admitted normal history passes. For each accepted history \(h\), some plan \(T_\nu\) contains \(E(h)\). Positivity of the weights gives
\[
\ell(h,\omega)\le\sum_{e\in T_\nu\setminus T_\omega}w(e)
\le D_{\mathcal F}.
\]
Together with the replay lower bound this proves equality. This reference monitor may retain a much richer history representation than a counter. The argument does not establish attainability of a restricted-pair lower bound when full pairwise replay fails.

\subsection{Optimal sparse quota (Theorem~\ref{thm:minimax})}

\paragraph{Normal completion}
An admitted plan \(T_S\) has \(|S|\le q\) replacements. Every prefix of every admitted order uses at most this many. The added replacement count therefore excludes none of its settlements. The unchanged exact permissions, funds, deduplication, and timing checks already support that normal history from the fresh state.

\paragraph{Upper bound}
Fix the true plan \(T_S\) and an executable attack history \(h\). Define
\[
I=\{i\in S:A_i\in E(h)\},\qquad
J=\{j\notin S:B_j\in E(h)\}.
\]
Distinct effect identities make the sets above enumerate each wrong debit exactly once. The sets are disjoint, and \(|I|\le|S|\le q\). Exact-effect deduplication and the shared replacement counter imply \(|J|\le q\). Hence
\[
\ell(h,S)=\sum_{i\in I}a_i+\sum_{j\in J}b_j\le D_q.
\]
A history may also contain a correct variant at a represented position when only exact-effect deduplication is present; that variant contributes zero to this loss and cannot invalidate the bound. Additional position-level deduplication only removes such histories.

\paragraph{Matching lower bound and ordinary resources}
Take a maximizing disjoint pair \((I,J)\). Both \(T_I\) and \(T_J\) are admitted plans. Replaying a complete normal history of \(T_J\) while \(T_I\) is true incurs exactly
\[
\sum_{e\in T_J\setminus T_I}w(e)
=\sum_{i\in I}a_i+\sum_{j\in J}b_j=D_q.
\]
The replay is a normal history with one effect per position, so it respects either deduplication scheme and uses normal funds, counts, and timing. Lemma~\ref{lem:replay} forces every completion-preserving policy to admit it under the assumed information contract. Thus \(D_q=D_{\mathcal F}\) for this family, and the count policy attains the minimum.

For debit-only funding, the outlay of \(T_S\) is
\[
\sum_i a_i+\sum_{i\in S}(b_i-a_i).
\]
Maximizing over \(|S|\le q\) selects up to \(q\) positive increments, yielding \(C\) in \eqref{eq:funding}. All payment amounts are positive, so no prefix requires more funding than its complete plan. A common balance of \(C\) is sufficient and, without additional inflows, necessary to support every plan. No extra attack-only funding is used in the lower bound.

\subsection{Dynamic program and witness verification}
\label{sec:certificate-dp}

Normalize the nonnegative integer quota to \(q\leftarrow\min(q,n)\). Let \(F_i(u,v)\) be the largest objective value over disjoint sets chosen from the first \(i\) positions with exactly \(u\) members of \(I\) and \(v\) members of \(J\). Set \(F_0(0,0)=0\) and all other initial states to \(-\infty\). A position is unused, contributes its original amount, or contributes its replacement amount:
\begin{equation}
\label{eq:certificate-dp}
\begin{aligned}
F_i(u,v)=\max\{&F_{i-1}(u,v),\\
              &F_{i-1}(u-1,v)+a_i,\\
              &F_{i-1}(u,v-1)+b_i\}.
\end{aligned}
\end{equation}
Omit terms with negative indices or with \(u,v>q\). The recurrence is exact by partitioning feasible choices according to the role of position \(i\). Induction on \(i\) gives
\(D_q=\max_{0\le u,v\le q}F_n(u,v)\).
It requires \(O(n(q+1)^2)\) arithmetic operations. Two layers need \(O((q+1)^2)\) storage for values. Storing one maximizing predecessor per state uses \(O(n(q+1)^2)\) storage and recovers \(I,J\) by backtracking. These are arithmetic-operation and storage counts; integer bit costs depend on the encoded amounts.

The certificate consists of the quota, the value \(D_q\), and the maximizing sets \(I,J\), which identify the true plan \(T_I\) and replay plan \(T_J\). A verifier checks their disjointness and sizes, reconstructs their exact effects, and sums the directed difference. It checks feasibility against the same funds, resource state, and admitted schedule used by the execution contract. This establishes a lower-bound witness; recomputing the dynamic program establishes that no larger pair exists. The structural upper-bound proof then applies to every accepted history. For a target strictly below \(D_q\), the verified normal replay certifies infeasibility under the same information and completion requirements.

At \(q=0\), \(D_0=0\). At \(q=n\), each position can be assigned independently to its more expensive variant, giving \(D_n=\sum_i\max(a_i,b_i)\). In particular, \(n=1,q=1\) gives \(\max(a_1,b_1)\). For \(q=1,n\ge2\), positivity allows two distinct positions, and \(D_1=\max_{i\ne j}(a_i+b_j)\). The same algorithm handles unequal original and replacement amounts; the monetary-comparator formula below requires equal amounts within each position.

\subsection{Matched amount cap (Theorem~\ref{thm:amount})}

The normal plan \(T_Q\) requires replacement spend \(H_q\), and every other normal plan spends at most \(H_q\) on replacements. Thus \(H_q\) is the smallest feasible shared amount cap.

For an accepted history under \(T_S\), let \(I\subseteq S\) contain its wrong-original positions and \(J\subseteq[n]\setminus S\) its wrong-replacement positions. They satisfy \(I\cap J=\varnothing\), \(|I|\le q\), and \(w(J)\le H_q\). Correct effects contribute no loss. Conversely, any such pair can be realized by submitting only \(\{A_i:i\in I\}\cup\{B_j:j\in J\}\) while \(T_I\) is true. These requests use distinct positions, at most \(n\) payments, and at most \(W\) funds. They satisfy the exact permissions, deduplication, global caps, and replacement amount cap. The comparison's scheduling premise ensures that this subset is executable within the window, even when \(|J|>q\) makes it inconsistent with every normal plan. Consequently
\[
R_{\mathrm{money}}=
\max_{\substack{I\cap J=\varnothing,\ |I|\le q\\ w(J)\le H_q}}
\bigl(w(I)+w(J)\bigr).
\]

Extend \(I\) to size \(q\) by adding unused positions or moving positions from \(J\) to \(I\). This never decreases the objective or increases replacement spend. If \(I\ne Q\), choose \(x\in Q\setminus I\) and \(y\in I\setminus Q\); then \(w_x\ge w_y\). If \(x\in J\), exchange \(x\) and \(y\) between \(J\) and \(I\). The objective is unchanged, while replacement spend weakly decreases. Otherwise replace \(y\) by \(x\) in \(I\); the objective weakly increases and replacement spend is unchanged. Repetition yields \(I=Q\). Optimizing the remaining replacements is exactly the tail subset-sum defining \(K_q\). This proves \(R_{\mathrm{money}}=H_q+K_q\), with witness \(T_Q\) and any maximizing tail subset.

For the count policy, \eqref{eq:dq} selects at most \(\min(2q,n)\) distinct positions, each of weight \(w_i\). Taking \(I=Q\) and \(J\) to be the largest \(\min(q,n-q)\) remaining positions attains their largest possible sum. Hence
\[
\begin{aligned}
R_{\mathrm{count}}=D_q
&=\operatorname{Top}_{\min(2q,n)}((w_i)_i)\\
&=H_q+S_q.
\end{aligned}
\]
Since \(S_q\le H_q\), its defining subset is feasible for \(K_q\), and \(K_q\ge S_q\). This proves the strict-improvement condition. When \(n\le2q\), \(S_q\) includes the whole tail, so \(K_q=S_q=W-H_q\). With all \(w_i\) equal, a tail subset fitting \(H_q\) has at most \(q\) elements, again giving equality. For \(q=0\) both risks are zero.

The two witness arguments have distinct roles. The count-policy optimum is forced by a complete normal replay under the information contract. The monetary optimum can use a mixed history with too many replacements to be normal. Its exact risk therefore also requires the stated feasibility of mixed subsets; funding sufficiency alone does not prove deadline feasibility. Extra sequence or context guards can reduce that optimum. The upper bounds on both policies continue to hold because they were derived from necessary conditions on accepted histories.

\subsection{Sharp relative reduction (Corollary~\ref{cor:sharp-gap})}

For \(n>2q\ge2\), the average of the largest \(q\) tail amounts is at least the average of all \(n-q\) tail amounts. Therefore
\[
S_q\ge\frac{q}{n-q}\sum_{i\notin Q}w_i
\ge\frac{q}{n-q}K_q.
\]
Together with \(K_q\le H_q\), this gives
\[
\begin{aligned}
\frac{K_q-S_q}{H_q+K_q}
&\le\left(1-\frac{q}{n-q}\right)\frac{K_q}{H_q+K_q}\\
&\le\frac{n-2q}{2(n-q)}.
\end{aligned}
\]
For \(q\) top amounts each equal to \(n-q\), and \(n-q\) tail amounts each equal to \(q\), we have \(H_q=K_q=q(n-q)\) and \(S_q=q^2\). Both inequalities are equalities, proving sharpness.

\subsection{Risk with partial waiting (Proposition~\ref{prop:deadline-split})}
\label{sec:deadline-proof}

The common schedule confines early settlement to \(U\). After verification, the executor accepts only selected effects at the remaining positions, so all wrong principal comes from the early phase. The projection onto \(U\) of the full plan family is exactly the family of subsets of size at most \(r\). Each projected plan extends to a full normal plan by leaving all later positions unchanged. Every full normal plan fits the early quota and has a feasible verified late phase by assumption. The count upper bound of Theorem~\ref{thm:minimax} applies to the projected family. For the lower bound, take the two plans that differ on the disjoint heaviest sets in \(U\), with at most \(r\) positions in each set. Their early normal prefixes must be accepted before the trusted evidence can distinguish the plans. Replaying one prefix in the other plan attains the bound. Once evidence arrives, later checks cannot undo an already committed debit.

The amount cap must accommodate the \(r\) heaviest early replacements in a normal prefix. Early effects using at most one variant per position fit the shared funds and successful-payment limit. When the mixed subsets can also meet the early deadlines, Theorem~\ref{thm:amount}'s exchange argument applies on \(U\), including the tight witness. If \(U\) is empty, both policies verify every payment and the risk is zero.

\subsection{Risk and exact effect-set representation}
\label{sec:representation}

This supplementary result concerns a specified representation of allowable effect sets. Let the target be the downward closure of \eqref{eq:envelope}: every subset of some admitted \(T_S\), with no additional order or context conditions. A logical group constraint has the form \(|H\cap G|\le c_G\), where \(H\) is the settled effect set, \(G\) any effect subset, and \(c_G\) a nonnegative integer. Initially only independent exact-effect deduplication is imposed. No position mutual exclusion or other history constraint is free in this representation comparison.

\begin{proposition}[Logical group constraints]
\label{prop:group-count}
For \(1\le q<n\), exactly \(n+1\) logical group constraints are necessary and sufficient to represent this downward closure. One shared replacement constraint already attains the optimal wrong-principal bound under Theorem~\ref{thm:minimax}'s resource and replay conditions. If position mutual exclusion is already enforced, that one replacement constraint also represents the downward closure.
\end{proposition}

\begin{proof}
The \(n\) constraints \(|H\cap\{A_i,B_i\}|\le1\), together with \(|H\cap\{B_1,\ldots,B_n\}|\le q\), are sufficient. Any passing set extends to a normal plan by keeping its replacement positions and filling unrepresented positions with originals.

For necessity, \(\{A_i,B_i\}\) is forbidden, while both singleton subsets are allowed. A valid group constraint excluding the pair must contain both effects and have cap one. For \(j\ne i\), \(\{A_i,A_j\}\) is contained in \(T_{\varnothing}\) and \(\{A_i,B_j\}\) in \(T_{\{j\}}\). The cap-one group therefore cannot include either variant from another position; it is exactly \(\{A_i,B_i\}\). Each of the \(n\) positions requires its own group. Those groups still allow \(q+1\) distinct replacements, so at least one further group is necessary. The stated construction reaches \(n+1\). The risk result follows from Theorem~\ref{thm:minimax}; with position mutual exclusion already present, only the replacement group remains necessary for exact representation.
\end{proof}

This counts logical inequalities in this representation. It does not compare storage bits, arbitrary state-machine encodings, or deployment cost. Other existing controls can alter the representation problem. A policy can attain the optimal risk while admitting an effect combination outside the normal family; the chosen loss function, rather than exact language equivalence, determines optimality.

\paragraph{Multiple amount caps can encode a count}
For every subset \(U\) of \(q+1\) replacement effects, impose an amount cap equal to the sum of the \(q\) largest amounts in \(U\). Any history using at most \(q\) replacements passes every cap. Any history using more than \(q\) replacements contains some such \(U\); positive amounts make its total exceed that cap. These overlapping monetary constraints therefore express the same count limit. When \(q=n\), no such constraints are needed. Thus the exact separation in Theorem~\ref{thm:amount} concerns one matched shared bucket and does not extend to unrestricted families of monetary constraints.

\section{Evaluation Details and Sensitivity}
\label{sec:evalmore}
\suppressfloats[t]

\subsection{Sources, Accounting, and Complete Grid}
\label{sec:eval-sources}

The AMLSim sample archives are \texttt{20K\_cycle200}, \texttt{20K\_fanin200}, and \texttt{20K\_fanin200cycle200}, containing 117,805, 118,250, and 120,558 rows, respectively. All three are fixed to revision \nolinkurl{7338a4bcb1af9bcfea2201ad7daccfe2a4d569ca}~\cite{amlSim}. The controlled sources are \texttt{controlled.jsonl} (160 cases, 820 actions) and \texttt{multistage\_exploratory.jsonl} (40 cases, 1,160 actions). The frozen protocols record their byte identities and the source order used for batching.

Every complete batch is scored at each quota for its width. Table~\ref{tab:batch-tails} distinguishes incomplete batches from the number of source rows they contain. These tails are excluded by width alone, and their statuses remain in the recorded results. No amount or completed batch is selected according to its benefit. Table~\ref{tab:full-risk-grid} reports all 24 strata per corpus. A batch not listed as a strict improvement ties; none has increased risk under the count policy.

\begin{table}[htbp]
\centering
\caption{Incomplete final batches and excluded source rows, by width. Batching restarts at each source boundary.}
\label{tab:batch-tails}
\footnotesize
\setlength{\tabcolsep}{5pt}
\begin{tabular}{@{}r rr@{\hspace{2em}}rr@{}}
\toprule
& \multicolumn{2}{c}{AMLSim} & \multicolumn{2}{c}{Controlled catalog} \\
\cmidrule(lr){2-3}\cmidrule(l){4-5}
\(n\)& Tails & Rows & Tails & Rows \\
\midrule
2 & 1 & 1 & 0 & 0 \\
4 & 3 & 5 & 0 & 0 \\
8 & 3 & 13 & 1 & 4 \\
16 & 3 & 37 & 2 & 12 \\
32 & 3 & 37 & 2 & 28 \\
64 & 3 & 133 & 2 & 60 \\
\bottomrule
\end{tabular}
\end{table}

\begin{table}[t]
\centering
\caption{Full quota grid. Each corpus reports \(\Delta\) from \eqref{eq:aggregate-reduction} in percent and strict improvements divided by complete batches. All remaining batches tie.}
\label{tab:full-risk-grid}
\footnotesize
\setlength{\tabcolsep}{3pt}
\begin{tabular}{@{}rr rr rr@{}}
\toprule
&&\multicolumn{2}{c}{AMLSim}&\multicolumn{2}{c}{Controlled catalog}\\
\cmidrule(lr){3-4}\cmidrule(l){5-6}
\(n\)&\(q\)&\(\Delta\)&Strict / full&\(\Delta\)&Strict / full\\
\midrule
2 & 1 & 0.00 & 0 / 178,306 & 0.00 & 0 / 990 \\
 & 2 & 0.00 & 0 / 178,306 & 0.00 & 0 / 990 \\
4 & 1 & 8.20 & 46,706 / 89,152 & 2.88 & 125 / 495 \\
 & 2 & 0.00 & 0 / 89,152 & 0.00 & 0 / 495 \\
 & 4 & 0.00 & 0 / 89,152 & 0.00 & 0 / 495 \\
8 & 1 & 7.50 & 37,717 / 44,575 & 2.52 & 139 / 247 \\
 & 2 & 14.38 & 44,334 / 44,575 & 8.40 & 225 / 247 \\
 & 4 & 0.00 & 0 / 44,575 & 0.00 & 0 / 247 \\
 & 8 & 0.00 & 0 / 44,575 & 0.00 & 0 / 247 \\
16 & 1 & 4.96 & 21,589 / 22,286 & 1.51 & 89 / 123 \\
 & 2 & 9.58 & 22,286 / 22,286 & 5.34 & 120 / 123 \\
 & 4 & 16.23 & 22,286 / 22,286 & 10.12 & 121 / 123 \\
 & 8 & 0.00 & 0 / 22,286 & 0.00 & 0 / 123 \\
 & 16 & 0.00 & 0 / 22,286 & 0.00 & 0 / 123 \\
32 & 1 & 2.62 & 11,101 / 11,143 & 0.95 & 48 / 61 \\
 & 4 & 10.23 & 11,143 / 11,143 & 4.37 & 59 / 61 \\
 & 8 & 16.71 & 11,143 / 11,143 & 12.18 & 61 / 61 \\
 & 16 & 0.00 & 0 / 11,143 & 0.00 & 0 / 61 \\
 & 32 & 0.00 & 0 / 11,143 & 0.00 & 0 / 61 \\
64 & 1 & 1.32 & 5,567 / 5,570 & 0.58 & 22 / 30 \\
 & 8 & 10.65 & 5,570 / 5,570 & 4.58 & 29 / 30 \\
 & 16 & 16.88 & 5,570 / 5,570 & 12.52 & 30 / 30 \\
 & 32 & 0.00 & 0 / 5,570 & 0.00 & 0 / 30 \\
 & 64 & 0.00 & 0 / 5,570 & 0.00 & 0 / 30 \\
\bottomrule
\end{tabular}
\end{table}

Table~\ref{tab:risk-sums} supplies the numerator components and denominators for the main result. The two corpora use different units and are never combined into one risk sum. Different widths and quotas also reuse the same source amounts and are interpreted separately.

\begin{table}[t]
\centering
\caption{Risk sums for the main \(q=n/4\) comparison. AMLSim sums use integer hundredths of a simulated unit; controlled sums use the records' integer-cent units.}
\label{tab:risk-sums}
\footnotesize
\setlength{\tabcolsep}{2.5pt}
\begin{tabular}{@{}r rr rr@{}}
\toprule
& \multicolumn{2}{c}{AMLSim} & \multicolumn{2}{c}{Controlled catalog} \\
\cmidrule(lr){2-3}\cmidrule(l){4-5}
\(n\)& \(\sum R_{\rm money}\) & \(\sum R_{\rm count}\) & \(\sum R_{\rm money}\) & \(\sum R_{\rm count}\) \\
\midrule
4 & 7,250,907,459 & 6,656,283,038 & 11,082,100 & 10,762,600 \\
8 & 7,980,138,430 & 6,832,479,052 & 12,030,900 & 11,020,800 \\
16 & 8,279,880,192 & 6,936,109,419 & 12,405,100 & 11,149,200 \\
32 & 8,397,594,049 & 6,994,630,728 & 13,086,500 & 11,492,700 \\
64 & 8,452,518,877 & 7,025,339,190 & 12,917,600 & 11,299,700 \\
\bottomrule
\end{tabular}
\end{table}

For \(q=n/4\), leaving out one AMLSim archive at a time gives ranges of 8.18--8.23\%, 14.36--14.40\%, 16.21--16.25\%, 16.67--16.73\%, and 16.84--16.92\% at widths 4, 8, 16, 32, and 64. At the same widths, evaluating the two controlled sources separately gives 2.01\% and 4.13\%, 7.55\% and 9.58\%, 9.97\% and 10.35\%, 13.11\% and 10.74\%, and 13.62\% and 10.87\%, in source order. These are composition checks on the finite corpora. The weaker global-caps-only AMLSim ablation gives reductions of 30.17\%, 29.58\%, and 29.25\% at widths 16, 32, and 64 with \(q=n/4\); it omits the replacement-specific monetary guard and is not the main comparator.

\subsection{Recorded Exactness Checks}
\label{sec:eval-validation}

For the narrow public grid, the recorded exact-risk calculations agree with 58 direct enumeration checks and with the earlier \(q=1\) receipt on all 334,319 narrow-width batches. The wide-grid validation records 24 independent meet-in-the-middle optima and 15 first-batch witnesses attaining a separately computed upper bound. The controlled catalog records independent meet-in-the-middle checks for the first complete batch at each width through 16, covering four batches across their quota grids. Its larger widths use the same exact integer recurrence and retain reconstructed witnesses.

The compiler validation separately reports 216 heterogeneous-amount instances and 72,816 candidate attack sets, as well as 39 equal-amount widths. It checks equality between the quota upper bound, the dynamic-programming certificate, and exhaustive normal-plan replay. The compiler replay receipt matches the 54 admissible public-structure cases and 760 funded original paths described below. These finite checks complement the proofs; they do not replace them.

\subsection{Original Controlled Paths}
\label{sec:original-paths}

This sensitivity keeps all 780 root-to-leaf paths from the 200 controlled source cases. Each path defines ordered batch positions with its original amounts, original payees, and available funding. Replacements and their exact rights are constructed, and the comparison evaluates the specified amendment family on each fixed path. It does not repartition actions into the new catalog batches or establish fidelity to additional source-workflow guards. There are 760 funded paths and 20 funding-infeasible paths; infeasible paths remain in the receipts and are not assigned zero risk.

At \(q=1\), the tight shared replacement spending cap has aggregate risk 15,571,200 cents, compared with 15,151,600 cents for the shared count, a reduction of 2.69\%. Of the 760 funded paths, 286 improve and 474 tie. The earlier 46.18\% result uses the weaker global-caps-only comparator on those paths, a different comparison from the same-scope monetary result. Table~\ref{tab:path-risk-grid} preserves every tested width and quota. All 20 funding-infeasible paths have width two. At \((n,q)=(8,2)\), the largest observed reduction is 8.49\%, with 39 strict improvements and one tie among 40 funded paths. Paths from the same source case are related and are not treated as independent samples.

\begin{table}[t]
\centering
\caption{Original controlled paths. Pairs give quota and aggregate risk reduction (\%) against the tight shared replacement spending cap.}
\label{tab:path-risk-grid}
\footnotesize
\setlength{\tabcolsep}{4pt}
\begin{tabular}{@{}rrl@{}}
\toprule
\(n\)& Funded paths & \((q,\Delta)\) \\
\midrule
2 & 60 & \((1,0.00),\;(2,0.00)\) \\
3 & 40 & \((1,6.27),\;(3,0.00)\) \\
4 & 100 & \((1,4.67),\;(2,0.00),\;(4,0.00)\) \\
5 & 160 & \((1,1.74),\;(2,5.99),\;(5,0.00)\) \\
6 & 320 & \((1,3.79),\;(3,0.00),\;(6,0.00)\) \\
7 & 40 & \((1,1.53),\;(3,7.73),\;(7,0.00)\) \\
8 & 40 & \((1,3.22),\;(2,8.49),\;(4,0.00),\;(8,0.00)\) \\
\bottomrule
\end{tabular}
\end{table}

\subsection{Public Authorization-Structure Sensitivity}
\label{sec:structure-sensitivity}

An earlier exploratory adaptation combines public Workflow Satisfiability Problem authorization entries~\cite{wspDataset} with amounts from the CRM Agent Benchmark~\cite{crmAgentBench}. It constructs recipients and singleton revisions that may change both payee and amount. All 1,476 instances pass the funding check, and 54 instances from 16 source pools admit every normal plan under the full available authorization graph. The remaining 1,422 fail authorization matching. The original workflow constraints are omitted, and the low admission rate limits the scope of the result.

On those 54 instances, aggregate wrong-payment principal falls from 4,560,950 cents under the optimal static and global-count controls to 3,107,000 cents under the shared count, a 31.88\% reduction. For the control with both global count and global spend, a certified lower bound of 4,168,250 cents yields a conservative reduction of at least 25.46\%. Table~\ref{tab:structure-risk} retains the width strata. All these comparisons use global controls; they do not establish the same-scope monetary comparison of Table~\ref{tab:quarter-risk}. The CRM ledger replay records 368 normal and 368 attack histories, with 2,960 transfers, exercising debit semantics. Stateful quota enforcement is tested separately through the MetaMask entry point.

\begin{table}[t]
\centering
\caption{Public-structure supplement: finite admitted subsets and reductions (\%) against global controls. The last column is a conservative reduction from a lower bound on the global count-and-spend optimum.}
\label{tab:structure-risk}
\footnotesize
\setlength{\tabcolsep}{4pt}
\begin{tabular}{@{}rrrr@{}}
\toprule
\(n\)& Admitted / all & Static or count & Count + spend \\
\midrule
4 & 20 / 648 & 25.86 & \(\ge 2.06\) \\
6 & 19 / 468 & 34.15 & \(\ge 32.95\) \\
8 & 15 / 360 & 35.38 & \(\ge 35.01\) \\
\bottomrule
\end{tabular}
\end{table}

\subsection{Native Fixture and Artifact Boundary}
\label{sec:eval-artifact}

The native result is pinned to MetaMask Delegation Framework commit \nolinkurl{b4b8c41fc912455a1af2300bfcd1276676b92a2e}. The fixture takes source rows one through eight of \texttt{20K\_cycle200}. In position order, the test-wei amounts are 14,780; 10,406; 18,377; 10,394; 12,894; 25,316; 22,583; and 10,483. Thus \(W=125{,}233\), \(H_2=47{,}899\), and the wrong-replacement sums in Table~\ref{tab:native-witness} can be checked directly. The amount witness uses six wrong payments; the count witness uses four. Both respect the common cap of eight payments and available funding.

The local execution receipt reports success for four tests: completion of all plans, rejected-branch rollback, the count-policy worst-debit witness, and the amount-policy worst-debit witness. Normal execution enumerates revision subsets in source-position order. The receipt reports no live wallets, production payments, deployment, or chain fork. It provides conformance evidence for this fixture, rather than measurements of production latency, attack incidence, recovery, or net economic loss.

\end{document}